\documentclass[10pt,preprint2,a4paper]{aastex}

\usepackage{amsmath}                
\usepackage{amsfonts}               
\usepackage{amssymb}                
\usepackage{epsfig}                 
\usepackage[left=2.1cm,top=2cm,right=1.2cm,nohead,nofoot]{geometry}

\newcommand{\cm}{{~\rm cm}}
\newcommand{\s}{{~\rm s}}
\newcommand{\km}{{~\rm km}}
\newcommand{\g}{{~\rm g}}
\newcommand{\K}{{~\rm K}}
\newcommand{\erg}{{~\rm erg}}
\newcommand{\yr}{{~\rm yr}}

\newcommand{\kpc}{{~\rm kpc}}
\newcommand{\Mpc}{{~\rm Mpc}}

\newcommand{\AU}{{~\rm AU}}

\newcommand{\days}{{~\rm days}}

\title{Common Powering Mechanism of Intermediate Luminosity Optical Transients and Luminous Blue Variables}

\author{Amit Kashi\altaffilmark{1} and Noam Soker\altaffilmark{1}}

\altaffiltext{1}{Department of Physics, Technion $-$ Israel Institute of
Technology, Haifa 32000 Israel; kashia@physics.technion.ac.il;
soker@physics.technion.ac.il.}

\begin{document}

\begin{abstract}
\small
We study recent Intermediate Luminosity Optical Transients (ILOTs) and major
eruptions of Luminous Blue Variables (LBVs),
and strengthen claims for a similar mechanism powering both.
This process is a short duration release of gravitational energy in a binary system.
In some ILOTs a merger occurs and one of the stars does not survive the transient event, e.g., V838~Mon and V1309~Sco.
In some transient events a rapid and short mass transfer process takes place and the two stars survive the transient event, e.g.,
the Great Eruption of $\eta$ Carinae.
We study new ILOTs and reanalyze known ones in light of new observations and models.
We reach our conclusion by analyzing these ILOTs using the Energy-Time Diagram (ETD)
where we plot the total energy of the eruption against its eruption timescale.
ILOTs and major LBV eruptions occupy the Optical Transient Stripe (OTS) in the ETD.
The upper boundary of the stripe is explained by our proposed model where a main sequence (or a slightly off-main sequence) star
accretes at a very high rate ($\lesssim 1 ~\rm{M_{\odot} \yr^{-1}}$) from a companion.
We identify one LBV, NGC~3432~OT, with two eruptions; one with weak total energy and the other with large total energy.
It bridges the regions of the ILOTs and LBVs in the OTS.
We further study the ILOT M85~OT2006 and show that it cannot be a nova, even not an extreme one.
We build a model where ILOTs can become optically thin in a timescale
of few years and the inflated envelope collapses into an accretion disk around the star.
Such an ILOT will evolve blue-ward after few years.
\end{abstract}


\small

\section{INTRODUCTION}
\label{sec:introduction}

Erupting objects with luminosities between those of
novae and supernovae (SNe) are being discovered at an accelerated rate
(e.g., Kulkarni et al. 2007a,b; Berger et al. 2009; Smith et al. 2009, 2011; Kasliwal et al. 2010a; Mason et al. 2010; Pastorello et al. 2010).
There is not yet an accepted term for these type of objects.
We will refer to them as Intermediate Luminosity Optical Transients (ILOTs).
Another popular name for these objects is luminous red novae (Kulkarni et al. 2007a,b), although using this nomenclature may be confusing,
as these transients are most probably not any kind of novae.

While the nature of these eruptions is poorly understood, there are several different proposed explanations
in the literature (see Kashi, Frankowski \& Soker 2010, hereafter KFS10, for further discussion).
KFS10 noticed that when rescaling the time axis for the V-band light curves
of ILOTs and major Luminous Blue Variable (LBV) eruptions, which at first seemed unconnected,
the shape becomes similar for a decline in more than $3$ magnitudes.
KFS10 suggested that these transients may have a similar powering mechanism --
mass accretion onto a main sequence (MS) star in a binary system.
KFS10 also showed that the light curves are different than the light curves of novae and SNe (type Ia and II).
However, the unique shape of the light curve of ILOTs is not yet understood.

By `major LBV eruptions' we refer to eruptions in which the luminosity rapidly ($\lesssim 1$ month) increases by a few magnitudes,
as opposed to less dramatic eruptions, such as S~Dor phases, weak eruptions in which
the luminosity is changed by $\lesssim 0.5$ magnitude in the V-band, or slow raises in magnitude.
All these processes are most probably internally related to the LBVs and are not to accretion processes which we discuss here.

Our proposed model of an eccentric binary interaction for ILOTs is directly based on the mergeburst model for V838~Mon
(Soker \& Tylenda 2003, 2006, 2007; Tylena \& Soker 2006), in which a $\sim 0.3~\rm{M_{\odot}}$ star
that merged with an $\sim 8~\rm{M_{\odot}}$ star created the outburst.
Indirectly, the binary model for ILOTs is supported by the similarities between ILOTS and LBVs (KFS10),
such as the massive binary system $\eta$ Car.
Damineli (1996) and Frew (2004; updated by Smith \& Frew 2011) noted that the beginning of the nineteenth century
major eruptions of $\eta$ Car occurred near periastron passages of the binary system.
A quantitative model for the periastron triggering of the $\eta$ Car nineteenth century major eruptions,
including mass loss and mass transfer (accretion) was recently conducted by Kashi \& Soker (2010).
Kashi (2010) further argued that the eruptions of the LBV P~Cygni in the seventeenth century were also triggered by
a periastron passages of an invisible companion.

The process of accretion onto a MS star was proposed in the past as an explanation
for several different types of objects, e.g., symbiotic stars (Kenyon \& Webbink 1984).
Kenyon \& Gallagher (1985) suggested that epochs of rapid mass exchange can account for the variability of some massive stars
although we note that $\eta$ Car and other LBVs are considered to be single stars in their study.
We instead, suggest that major LBV eruptions are triggered by interaction with a close companion to the very massive star
(Kashi \& Soker 2010; Kashi 2010).
In this process, accretion of LBV material onto a MS (or slightly off-MS) star takes place in some cases in the form of an accretion disk.
Bath \& Pringle (1981) discussed eruptions from accretion episodes triggered by an accretion disk instability
onto a MS star and a white dwarf (WD) in the context of dwarf novae.
Though the light curve quite nicely matches the light curve of ILOTs when rescaling the time scales,
the physical mechanism of ILOTs is probably different, as there is little chance for observing the temporary accretion disk
through the inflated envelope.

Smith (2011) also suggested that some (but not all) ILOTs are related to binary interaction
and attributed the erupting extra energy to the motion of the companion through the primary's envelope.
It is however unclear how in such a scenario the energy of the eruption can be accounted for
as there is too little orbital energy.

In this paper we extend our model for ILOTs to account for recent observations.
In section \ref{sec:theEnergyTimeDiagram} we describe the Energy-Time Diagram (ETD).
In section \ref{sec:Recent_Transients} we implement it for recent transients, demonstrating it to be a powerful tool.
In section \ref{sec:Known_Transients} we explain why recent claims for M85~OT2006 being a nova are problematic,
and suggest that the blue spectra observed in late times for M31~RV comes from an accretion disk around the progenitor,
which is surrounded by an optically thin shell.
We summarize in section \ref{sec:summary}.

\section{THE ENERGY-TIME DIAGRAM}
\label{sec:theEnergyTimeDiagram}

\begin{figure*}
\resizebox{1\textwidth}{!}{\includegraphics{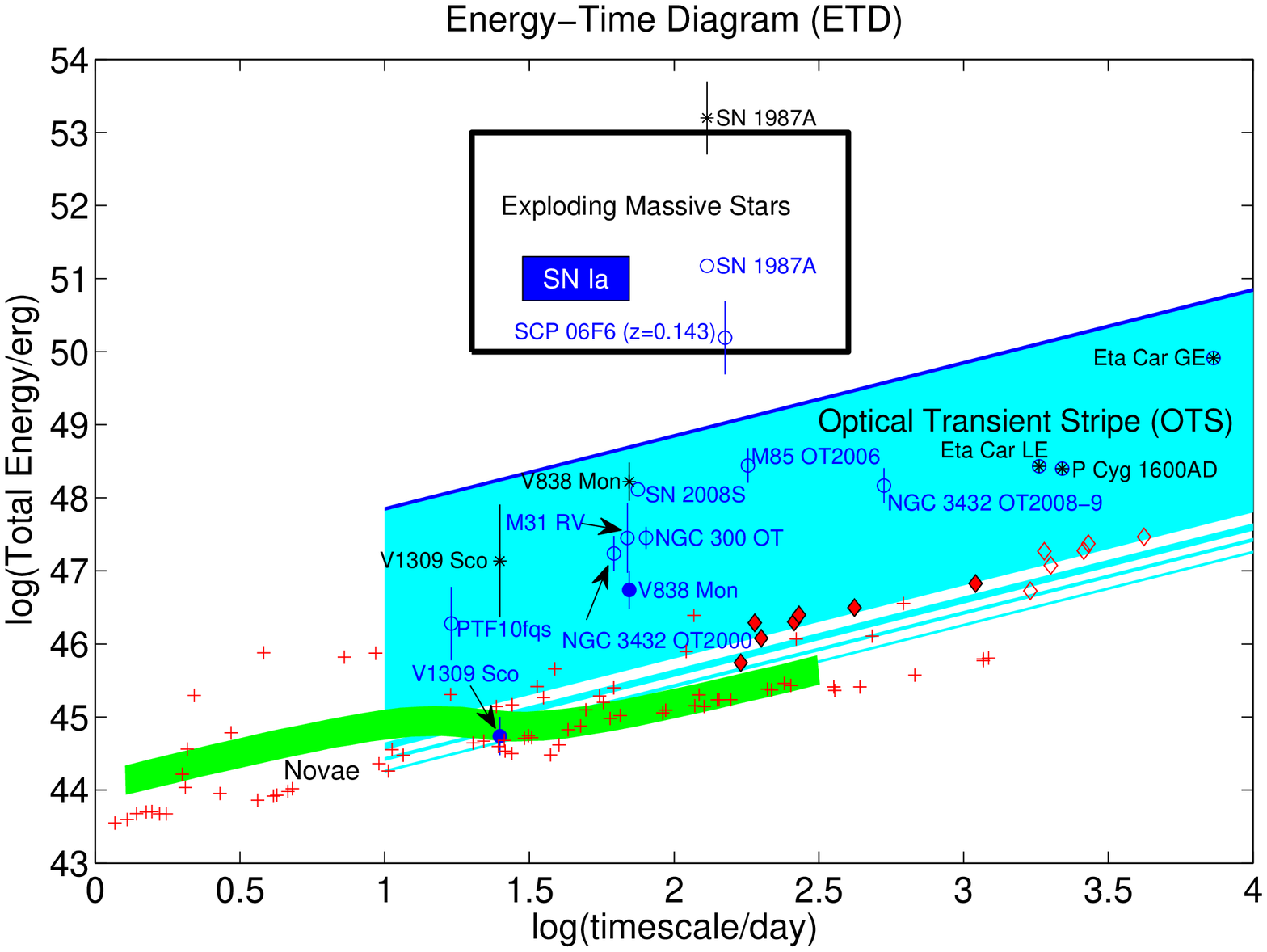}}
\caption{\footnotesize{The Energy-Time Diagram (ETD).
Blue empty circles represent the total (radiated plus kinetic) energy of the transients $E_{\rm{tot}}$
as a function of the duration $t$ of their eruptions
(generally defined as the time it took the transient to decrease in 3 magnitudes in the V-band; see exceptions in the text).
Blue filled circles indicate that the transient is a mergeburst.
The total energy does not include the energy which is deposited in lifting the envelope that does not escape from the star.
Where a model is available, we used is to calculate the total gravitational energy released by the accreted mass: the available energy.
This is marked by a black asterisk.
For LBV eruptions this energy is equal to the radiated plus kinetic energy, as there is no inflated envelope (hence the symbols overlap).
In some cases, such as in V838~Mon, the total released gravitational energy is considerably larger than the combined radiated and kinetic energy forms.
Four new eruptions that were added to the diagram of KFS10 are listed in Table \ref{tab:data},
as well as updated values for eruptions which appear in that paper.
Green wavy band: Nova models computed using luminosity and duration from Kulkarni et al. (2007a).
Nova models from Yaron et al. (2005) are marked with red crosses.
Unlike in KFS10 where we excluded part of the nova models of Yaron et al. (2005), here we present all these models.
New models from Shara et al. (2010a) are also plotted (red open diamonds and red filled diamonds; see text for details).
Eta Car GE and LE are the Great and Lesser Eruptions of $\eta$ Car, respectively.
The shaded area is the Optical Transient Stripe (OTS), where the eruptions we try to unify under the same energy source are located.
The thick blue line is the upper limit of the OTS.
This upper limit has a more or less constant luminosity as derived in equation (\ref{eq:Lmax}).
We conclude that most of these transients are associated with mass-transfer processes onto a MS star
(we consider a merger process to be an extreme case of mass transfer in the context of our model).
The LBV eruptions NGC~3432~OT bridge the upper and lower parts of the OTS.}}
\label{fig:totEvst}
\end{figure*}

The ETD (Figure \ref{fig:totEvst}) presents the total energy of the transients, radiated plus kinetic,
as a function of the duration of their eruptions, defined as a drop of $3$ magnitudes in the V-band.
Other energy sinks, such as the energy required to inflate the envelope,
are not included, as they are not observed.
However, in some cases the omitted energy may be considerable, as for the case of V838~Mon,
where the energy required to inflate the envelope is more than $10$
times larger than the value we present in Figure \ref{fig:totEvst} by the blue filled circle (Tylenda \& Soker 2006).
To account for the total \emph{available} energy we calculate and present the gravitational
energy released by the accreted mass (marked by a black asterisk in the ETD).
The advantage is that the total available energy better reflects the physical processes behind the transient,
as according to our model these transient events are powered by gravitational energy.
The disadvantage is that the total energy comes from a model for each transient event,
and models exist for only a fraction of all transient events.
For LBV eruptions this energy is equal to the radiated plus kinetic energy, as there is no inflated envelope.
We estimate the total available energy for some ILOTs as we later explain.
However, as stated, for most ILOTs the observations and models are not yet detailed enough to perform this estimate,
and we can only present the estimated radiated plus kinetic energy.
For a pedagogical purpose and comparison, we also present in Fig \ref{fig:totEvst} the available energy for the Type II supernova 1987A.
Here the available energy comes from a collapsing core.
Most of the energy is carried by neutrinos and anti-neutrinos.
Only $\sim 1 \%$ of the energy goes into kinetic energy and radiation, as marked.

Many transients occupy the Optical Transients Stripe (OTS) in the ETD, extending over $\sim3$ orders of magnitude.
The upper-right region of the OTS is occupied by major LBV eruptions,
while the lower-left region is occupied by ILOTs.
Figure \ref{fig:totEvst} presents the updated ETD, with recent optical transients added and other optical transient events at their updated locations,
according to the values in Table \ref{tab:data}.
\begin{table*}
\caption{\footnotesize{
Data used to update the Energy-Time Diagram (ETD; Figure \ref{fig:totEvst}).
For the other objects see table 1 in KFS10.
}}
\label{tab:data} \medskip
\footnotesize{
\begin{tabular}{l l l p{2.5cm} p{2cm} p{3.5cm}}
\hline
\multicolumn{1}{l}{Transient} & \multicolumn{1}{l}{Duration} & \multicolumn{1}{l}{Total energy}         & \multicolumn{1}{l}{Total available energy}  & \multicolumn{1}{l}{Reference} & Comments \\
\multicolumn{1}{l}{}          & \multicolumn{1}{l}{}         & \multicolumn{1}{l}{Radiated and Kinetic} & \multicolumn{1}{l}{incl. infl. envelope}    & \multicolumn{1}{l}{}          &  \\
\multicolumn{1}{l}{}          & \multicolumn{1}{l}{[days]}   & \multicolumn{1}{l}{[$10^{48} \erg$]}     & \multicolumn{1}{l}{[$10^{48} \erg$]}        & \multicolumn{1}{l}{}          & \multicolumn{1}{l}{}\\
\hline
&  &  &  &  &\\
V1309~Sco              & $25$    & ($5$~--~$15$)$\times 10^{-4}$ & $0.023$~--~$0.79$               & Mason et al. (2010)                                                       & a. Assuming a distance of $8 \kpc$. b. Assuming the same ratio between total and radiated energy as for V838 Mon.\\
PTF10fqs               & $>17$   & $0.03$~--~$0.3$               & no model available              & Kasliwal et al. (2010a); Kasliwal 2010 priv. comm.                        & Temporary result based on R-band observations. \\
NGC~3432~OT2000        & $62$    & $0.1$~--~$0.3$                & no model available              & Wagner et al. (2004)                                                      & Light curves was corrected for bolometric luminosity using blackbody temperature. \\
NGC~3432~OT2008-9      & $531$   & $0.85$~--~$2.5$               & no model available              & Pastorello et al. (2010)                                                  & Light curves was corrected for bolometric luminosity using blackbody temperature. \\
V838~Mon (updated)     & $70$    & $0.03$~--~$0.1$               & $0.9$~--~$3$                    & Tylenda \& Soker (2006)                                                   & \\
M31~RV (updated)       & $70$    & $0.094$~--~$0.85$             & no model available              & Rich et al. (1989); Mould et al. (1990)                                   & The lower (upper) energy estimate is for $0.01~(0.1)~\rm{M_{\odot}}$ ejected in the eruption. \\
M85~OT2006 (updated)   & $180$   & $1.6$~--~$4.6$                & no model available              & This work; Kulkarni et al. (2007a); Ofek et al. (2008); Rau et al. (2007) & We estimate the ejected mass to be $0.2$~--~$0.6 \rm{M_{\odot}}$. \\
\hline
\end{tabular}
}
\end{table*}

As additional transients are added to the ETD, we find it necessary to update the location of the OTS.
The $1838$~--~$1858$ Great Eruption (GE) of $\eta$ Car is presently considered as the highest energy
and longest duration eruption in the OTS.
We note that the $1838$~--~$1858$ Great Eruption of $\eta$ Car is treated as one major eruption,
despite the fact that several peaks are observed throughout the duration of the outburst (KFS10).
Smith \& Frew (2011), on the other hand, treat each of the high rises in luminosity
during the Great Eruption of $\eta$ Car as a short duration transient event by itself.

Another object, SN 2008HA, seems to be too luminous to belong to the OTS following the new
less luminous transients we added (see figure 3 of KFS10).
We therefore conclude that it most likely does not belong to this group of transient events and remove it from the ETD.
It is important to note that the purpose of the OTS is not to account for peculiar types of SNe,
and as we focus on objects which have the properties of accretion powered transient events we do not include such objects in the ETD.

One can use a luminosity-time diagram (Kulkarni et al. 2007a)
to draw conclusions based on the observational properties of the objects.
But the usage of the ETD, with total energy instead of the luminosity, better
accounts for the physical properties of the different types of objects
and emphasize that they form a stripe (the OTS) that distinguish them from other objects.
The OTS hints to a common mechanism for powering transients
which we propose to be associated with a mass accretion onto a MS star, or a star just off the MS.
The mass transfer takes place in a binary system, although the companion might not survive the transient event.
The OTS, as presented in Figure \ref{fig:totEvst}, has a constant slope of 1 in the
$\log(E_{\rm{tot}}/\erg)$~--~$\log(t/\rm{day})$ plane,
although with the addition of more objects in the future the slope might change a little.
The lower part of the OTS overlaps with novae.
This should not be regarded as a problem for the following reasons:
(1) Only one object, V1309~Sco occupies this overlapping region, and this occurs only when the energy considered is less than the total available energy.
When all the available energy is considered, it is located at much higher energies and far from the overlapping region.
(2) The ILOTs have other common properties that distinguish them from novae (e.g., their light curves).

The lower part is not yet theoretically constrained, and should be regarded as an observational constraint (selection effects).
From above, the domain of OTS is limited by theoretical considerations.
Let $M_a$ and $R_a$ be the mass and radius of the star accreting the mass respectively (a MS or a slightly evolved star).
Star `$b$' is the one that supplies the mass to the accretion; it can be a MS (or a little off-MS) star that is completely destroyed,
as in the models for V838~Mon (Soker \& Tylenda 2003; Tylenda \& Soker 2006) and V1309~Sco (Tylenda et al. 2011),
or an evolved star in an unstable phase of evolution that loses a huge amount of mass,
as in the model for the Great Eruption of $\eta$ Car (Kashi \& Soker 2010).

The average total gravitational power is the average accretion rate times the potential well of the accreting star
\begin{equation}
L_G=\frac{G M_a \dot{M_a}}{R_a}.
\label{eq:L}
\end{equation}
In the binary model discussed here, accreted mass is likely to form an accretion disk or an accretion belt.
The accretion time must be longer than the viscosity time scale for the accreted mass to lose its angular momentum.
According to Dubus et al. (2001) the viscous timescale is
\begin{equation}
\begin{split}
t_{\rm{visc}} &\simeq \frac{R_a^2}{\nu}
\simeq 73
\left(\frac{\alpha}{0.1}\right)^{-1}
\left(\frac{H/R_a}{0.1}\right)^{-1}
\left(\frac{C_s/v_\phi}{0.1}\right)^{-1}\\
&\left(\frac{R_a}{5 \rm{R_{\odot}}}\right)^{3/2}
\left(\frac{M_a}{8 \rm{M_{\odot}}}\right)^{-1/2} \days, 
\end{split}
\label{eq:tvisc1}
\end{equation}
where $\nu$ is the viscosity of the disk, $H$ is the thickness of the disk,
$C_s$ is the sound speed and $v_\phi$ is the Keplerian velocity.
We scale $M_a$ and $R_a$ in equation (\ref{eq:tvisc1}) according to the parameters of V838~Mon (Tylenda 2005).
For these parameters the viscous to Keplerian times ratio is $\chi \equiv t_{\rm{visc}}/t_K \simeq 160$.

The accreted mass is determined by the details of the binary interaction process, and differs from object to object.
We scale it by $M_{\rm{acc}} = \eta_a M_a$.
Based on the modeled systems (V838~Mon, V~1309~Sco, $\eta$ Car) this mass fraction is $\eta_a \lesssim 0.1$ with a large variation.
The value of $\eta_a \lesssim 0.1$ can be understood as follows.
If the MS (or slightly off-MS) star collides with a star and tidally disrupts it, as in the model for V838~Mon (Soker \& Tylenda 2003; Tylenda \& Soker 2006),
the destructed star is likely to be of much lower mass than the accretor $M_{\rm{acc}} \lesssim M_b \lesssim 0.3 M_a$.
Another possibility is that an evolved star loses a huge amount of mass.
In that case it is possible that the accretor will gain only a small fraction of the ejected mass,
as in the scenario for the Great Eruption of $\eta$ Carinae (Kashi \& Soker 2010).
Here again we expect $M_{\rm{acc}} \lesssim 0.1 M_a$.

The viscous time scale gives an upper limit on the accretion rate
\begin{equation}
\begin{split}
\dot{M} < \frac{\eta_a M_a}{t_{\rm{visc}}}
&\simeq 4
\left(\frac{\eta_a}{0.1}\right)
\left(\frac{\alpha}{0.1}\right)
\left(\frac{H/R_a}{0.1}\right)
\left(\frac{C_s/v_\phi}{0.1}\right)\\
&\left(\frac{R_a}{5 \rm{R_{\odot}}}\right)^{-3/2}
\left(\frac{M_a}{8 \rm{M_{\odot}}}\right)^{3/2} ~\rm{M_{\odot} \yr^{-1}}. 
\end{split}
\label{eq:dotM}
\end{equation}
The maximum gravitational power is therefore
\begin{equation}
\begin{split}
L_G < L_{\rm{max}} = \frac{GM_a\dot{M_a}}{R_a}& \simeq 7.7 \times 10^{41}
\left(\frac{\eta_a}{0.1}\right)
\left(\frac{\chi}{160}\right)^{-1}\\
&\left(\frac{R_a}{5 \rm{R_{\odot}}}\right)^{5/2}
\left(\frac{M_a}{8 \rm{M_{\odot}}}\right)^{-5/2}
\erg ~\rm{s^{-1}},
\end{split}
\label{eq:Lmax}
\end{equation}
where we replaced the parameters of the viscous time scale with the ratio of viscous to Keplerian time $\chi$.
Equation (\ref{eq:Lmax}) sets the upper bound on the OTS in the ETD, plotted as blue line at the upper edge of the OTS.
We note that the location of this line may change if the accretion efficiency $\eta$ is different
and/or the stellar parameters of the accreting star are different.
For most of the ILOTs the accretion efficiency is lower, hence they are located below this line,
giving rise to the relatively large width of the OTS.
The uncertainty in $\eta_a$ is large and in extreme cases may be even $>1$.
Therefore, on rare occasions we expect to find objects slightly above the upper line drawn in the figure.

It seems that the ETD has the potential to serve as a good diagnostic for classifying and perhaps better understanding optical transients.
The next section demonstrates how we find recent transients to belong, or not to belong to the OTS
and elaborate on one object that bridges the upper LBV and lower ILOT part.

\section{RECENT OPTICAL TRANSIENTS}
\label{sec:Recent_Transients}

We add new objects to the original ETD of KFS10, which include new features that strengthen the ETD as an analyzing tool.
We discuss each object with the new features it brings.

\textbf{V1309~Sco:}
The variable star V1309 Sco was discovered on Sep 2.5, 2008 (Nakano 2008).
The distance to V1309~Sco was constrained to be $<8\kpc$ (Mason et al. 2010),
and later better estimated as $\sim 3\kpc$ (Tylenda et al. 2011).
Mason et al. (2010) considered it to be a candidate twin of V838~Mon.
Below we strengthen this claim for a similarity with V838~Mon.
Tylenda et al. (2011) later conducted a thorough study of the progenitor of V1309~Sco and
convincingly showed it to be a mergeburst similar to that of V838~Mon, and by that
strongly supported the early claim made by Mason et al. (2010) and us.

\begin{figure}[t]
\resizebox{0.5\textwidth}{!}{\includegraphics{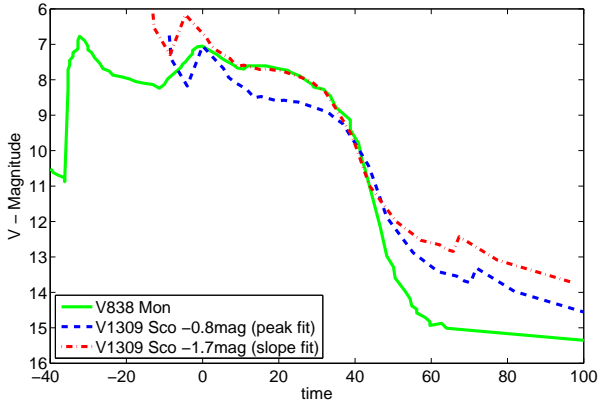}}
\caption{\footnotesize{Comparing the V-band light curves of V838~Mon (Sparks et al. 2008; solid-green line) and V1309~Sco (Mason et al. 2010).
We take the visible light curve of V1309~Sco (Mason et al. 2010) as complementary to the V-band light curve when there are no
observations from the latter, i.e., for the first small peak at the beginning of the eruption.
We shift and rescale the time of V1309~Sco but leave the magnitude scale unchanged (only shifted vertically).
Dashed-blue line: matching the last peak of V838~Mon with the second peak of V1309~Sco and then rescaling the time to match the light curves.
One unit of time corresponds to 1 day for V838~Mon, and $0.6$ days for V1309~Sco.
The light curve of V838~Mon has a curved slope in the $\sim 30$ days following its second
peak, before it becomes steeper.
Dot-dashed red line: matching the initial decline of the light curves (the `knee').
Note that the time scale is the same ($0.6$ days) for the two plots of V1309~Sco, but the peak magnitude and the time shift are different.
Also note that the eruption of V838~Mon begin approximately at $t=-36 \days$, and its beginning is not shown here.
Namely, the entire duration of the eruption of V838~Mon is longer than the eruption of V1309~Sco.
The resemblance between the two eruptions is much better by the dot-dashed red line,
which follow the light curve of V838~Mon for a decline of almost 4 magnitudes.}}
\label{fig:V1309Sco_V838Mon_DIM}
\end{figure}

In Figure \ref{fig:V1309Sco_V838Mon_DIM} we compare the V-band light curve of V1309~Sco
(from fig. 1 of Mason et al. 2010) with that of V838~Mon (from fig. 5 of Sparks et al. 2008).
We take the visible light curve of V1309~Sco (Mason et al. 2010) as complementary to the V-band light curve when there are no
observations from the latter, i.e., for the first small peak at the beginning of the eruption.
It is interesting to note that both objects show more than one peak at the onset of their eruption.
We shift and scale the time scale of V1309~Sco (by a scale factor of $0.6$) but leave the magnitude scale unchanged (only shifted vertically).
We find that the resemblance between the two eruptions is remarkable, as the time-scaled light curve of V1309~Sco
follow the light curve of V838 Mon for a decline of almost 4 magnitudes.

We find that the similarity between V1309~Sco and V838~Mon goes much further than the similarity in the shape of the light curves.
Where data are available, the I-band is $\sim 2$ magnitudes above the V-band in both objects.
Moreover, there are no high excitation lines in either object,
and for both objects the expansion velocities are much less than those of novae.

Using archive data, Tylenda et al. (2011) discovered that the progenitor
possessed a periodic behavior with a period of $\sim 1.4 \days$.
For the six pre-outburst years analyzed they found the period to decrease.
They found the progenitor to be consistent with a binary system with a total mass of $\sim 1$~--~$3~\rm{M_{\odot}}$,
and confirmed that the eruption was a mergeburst, similar to, but less energetic than V838~Mon (Soker \& Tylenda 2006, 2007).
Integrating over time the bolometric luminosity from Tylenda et al. (2011) we find that the
bolometric radiated energy is $E_{\rm{rad}}$(V1309~Sco) $\simeq 4 \times 10^{44} \erg$. 
This is $\sim 0.017$ times the radiated energy in V838~Mon (Tylenda 2005).
Since the bolometric radiated energy ratio between V1309~Sco and V838~Mon is $\sim 0.02$ and both are mergebursts,
we take the total energy of V1309~Sco to also be $\sim 0.02$ of the total energy of V838~Mon.
This gives $E_{\rm{tot}}$(V1309~Sco) $\simeq 3$~--~$10 \times 10^{44} \erg$ (but see the total available gravitational energy estimate below).
Following this assumption we position V1309~Sco on the ETD (Figure \ref{fig:totEvst}).

We note that in V838~Mon most of the energy goes into inflating the huge envelope (Tylenda \& Soker 2006).
We therefore calculate for both transients the total available energy, which is the gravitational energy of the merger, is
\begin{equation}
\begin{split}
E_{\rm{tot,ava}} &\simeq \frac{1}{2}\frac{G M_a \Delta M_{\rm{acc}}}{R_a} \\
&\simeq 1.9 \times 10^{48}
\left(\frac{M_a}{\rm{M_{\odot}}}\right)
\left(\frac{\Delta M_{\rm{acc}}}{\rm{M_{\odot}}}\right)
\left(\frac{R_a}{\rm{R_{\odot}}}\right)^{-1} \erg, 
\end{split}
\label{eq:total_available_energy}
\end{equation}
where $M_a$ is the mass of the (more massive and dense) primary accretor, $\Delta M_{\rm{acc}}$ is the accreted mass of the companion,
and $R_a$ is the radius of accretor.
We find that for V838~Mon the total available energy
is $E_{\rm{tot,ava}}$(V838~Mon) $\simeq 9$~--~$30 \times 10^{47} \erg$, 
taking $(M_a,\Delta M_{\rm{acc}},R_a)=(8,0.3,5)$ in solar units for the lower estimate
and a factor of 3.3 for the higher estimate.
For V1309~Sco we get $E_{\rm{tot,ava}}$(V1309~Sco) $\simeq 2.3$~--~$79 \times 10^{46} \erg$, 
taking $(M_a,\Delta M_{\rm{acc}},R_a)=(0.93,0.07,5.4)$ in solar units for the lower estimate,
and (1.5,1.5,5.4) for the upper estimate, based on the results of Tylenda et al. (2011).
As noted in section \ref{sec:theEnergyTimeDiagram}, the total available energy is a better measure of the physical
process behind the accretion powered ILOTs.

\textbf{PTF10fqs:}
The optical transient PTF10fqs was discovered in a spiral arm of M99
by the Palomar Transient Factory (PTF) in Apr 16, 2010, and showed
eruption characteristics similar to those of M85~OT2006, SN 2008S, and NGC~300OT (Kasliwal et al. 2010a).
Taking the R-band magnitude (M. Kasliwal, private communication) and the
spectra given in figure 10 of Kasliwal et al. (2010a)
we find that the R-band luminosity should be corrected by a factor of $\sim 1.1$ to obtain the bolometric luminosity.
We integrate the bolometric luminosity over the first $\sim 30$ days after detection to obtain the total radiated energy
$E_{\rm{rad}}$(PTF10fqs) $\simeq 3 \times 10^{46} \erg$.
The total energy is estimated to be up to $10$ times larger, about the maximum ratio observed for ILOTs.
The peak luminosity occurred $\sim 13$ days after detection and therefore we have data for only $\sim 17$ days for the decline,
during which the R-band decreased by only $\sim 0.4$ magnitudes.
We note that the decline in the V-band was steeper than in the R-band.
We find PTF10fqs to be located on the lower edge of the OTS
However, this placement is temporary and once more observations are
available we will be able to deduce the total energy and decline timescale.

\textbf{NGC~3432~OT:}
A series of LBV eruptions were observed from an LBV in the spiral galaxy NGC~3432 (Pastorello et al. 2010).
The first in the series started in May 3, 2000 (NGC~3432~OT2000; also referred to as SN~2000ch) and lasted 62 days (Wagner et al. 2004).
{}From Oct 7, 2008, a series of three eruptions (NGC~3432~OT2008-9) has been observed from the same star (Pastorello et al. 2010),
lasting a total of $\sim 531$ days.
We refer to the above durations as the decline times in the ETD as they are the characteristic times of the transients
(a similar treatment was done for the other LBVs in KFS10).
We consider the 2000 eruption as a separate one from those of 2008-9 because there was a long quiescence period
between them which had on average $\sim 1.5$~--~$2$ weaker V-band magnitude than at the times of the eruptions.
All four peaks were followed by a temporary decrease in luminosity, possibly caused by a dust shell which later
dissipated, and allowed the remainder of the decline in the eruption luminosity to be observed (Pastorello et al. 2010).
Observing the spectra, Wagner et al. (2004) derived that the effective temperature of the star during NGC~3432~OT2000 was $\sim 7800\K$,
and for NGC~3432~OT2008-9 Pastorello et al. (2010) obtained $\sim 9000\K$.
More spectra given by Smith et al. (2011) approximately 4 years after the first outburst,
shows Balmer lines with P Cyg profiles, blueshifted in their centeroids.
The FWHM was $\sim 1500 \km \s^{-1}$ with wide wings up to $-2500 \km \s^{-1}$ and $4300 \km \s^{-1}$ in the blue and red, respectively
(Smith et al. 2011).
This is an indication of absorbing material which was probably ejected during the LBV's major eruption.

We corrected the light curves of NGC~3432~OT for bolometric luminosities using blackbody temperatures
for a star with the temperatures given above, and integrated the light curves.
We did this procedure for the R, V, and B bands and derived the approximatly radiated energies of
$E_{\rm{rad}}$(NGC~3432~OT2000) $\simeq 10^{47} \erg$, and
$E_{\rm{rad}}$(NGC~3432~OT2008-9) $\simeq 8.5 \times 10^{47} \erg$.
As these transients are major LBV eruptions we estimate the total energies to be $\sim 3$ times larger than the radiated energies,
same as for the $\eta$ Car Great eruption (KFS10).

NGC~3423~OT overcomes a drawback of our early analysis in KFS10.
NGC~3423~OT2008-9 is located in the central part of the OTS, namely in
the $2.3 \lesssim \log(t/\days) \lesssim 3.3$ interval between the upper part
associated with LBVs and the lower part associated with ILOTs.
Also, NGC~3423~OT2000 shows low energy and short timescale which position it at the heart of the ILOTs part of the OTS,
though it is actually a major LBV eruption.
NGC~3423~OT therefore bridges the two parts of the OTS, namely the ILOTs and the major LBV eruptions.
This a strong support to the suggestion of KFS10 that LBV major eruptions and ILOTs at the lower part of the OTS
have a similar powering mechanism.

\textbf{SN~2002bu:}
Smith et al. (2011) presented new observations as well as archival study of various transients.
Though most of them do not fall in the category of ILOTs there is one relevant here.

SN~2002bu erupted in NGC~4242 (Puckett \& Gauthier 2002) and was initially thought to be a type II-P supernova.
The V magnitude declined by 2.67 magnitude in $\sim 84$ days (Smith et al. 2011).
The spectra given by Smith et al. (2011) shows that it reddened during the 81 days following first discovery,
supposedly by a dust shell forming around it when the ejected material has cooled down to grain forming temperatures.
This is in contrast to typical SN spectra evolution, leading Smith et al. (2011) to suggest that it is not a SN,
but rather a ``SN impostor''.
The ejection velocity, obtained from the H$\alpha$ profile was $\sim 900 \km \s^{-1}$,
but the ejected mass is hard to estimate.

Like for V1309~Sco, we compare the V-band light curve of SN~2002bu
from table 3 of Smith et al. (2011) to that of V838~Mon,
and find it to follow its decline quite well when rescaling its time axis by a factor of 2.

The distance estimates to NGC~4242 are not very accurate and we adopt the estimate of $\sim 8 \Mpc$
(Thompson et al. 2009).
accordingly, the peak luminosity translates to $\sim 8.8 \times 10^{40} \erg \s^{-1}$.
We integrated the luminosity from the time of discovery ($\sim 5$ days before peak luminosity)
to the last available observation ($\sim 84$ days after the peak luminosity).
Using bolometric correction to find the radiated energy, as we did for the ILOT in NGC~3432 has proved
difficult as there was no conclusive blackbody temperature to associate with the emission.
However using approximate values obtained from the shape of the spectra we find the radiated energy to be
$E_{\rm{rad}}$(SN~2002bu) $\simeq 3 \times 10^{48} \erg$.
With no mass estimate to calculate the kinetic energy,
we can only speculate that the total energy is in the order of $\sim 10^{49-50} \erg$,
larger than that of other ILOTs with the same time scale.

As SN~2002bu is far out of the OTS, we conclude that it is not an ILOT, and the most likely explanation is that it is a peculiar class of SN.
This demonstrates that the ETD can also be used to identify transients with different physical properties than the ones considered as
accretion powered ILOTs.

\section{REEXAMINING KNOWN TRANSIENTS}
\label{sec:Known_Transients}

\subsection{Why M85~OT2006 Cannot be a Nova}
\label{subsec:Known_Transients_M85}

The transient M85~OT2006 was discovered in the lenticular galaxy M85 in Jan 7, 2006 (Kulkarni et al. 2007a).
It had a peak luminosity of $\sim 2 \times 10^{40} \erg \s^{-1}$ and total radiated energy of
$E_{\rm{rad}}$(M85~OT2006) $\simeq 6 \times 10^{46} \erg$ over
a duration of $\sim 180 \days$ (Kulkarni et al. 2007a).
The progenitor's mass was estimated to be $< 7~\rm{M_{\odot}}$ (Ofek et al. 2008)
and it was suggested that the origin of M85~OT2006 is a stellar merger (Kulkarni et al. 2007a).

Rau et al. (2007) estimated that the effective temperature and the stellar radius at the time of peak luminosity
were $T_{\rm{eff}} \simeq 4600\K$ and $R \simeq 3600~\rm{R_{\odot}}$ respectively.
At a later time in the eruption the star cooled down and expanded to have
$T_{\rm{eff}} \simeq 950\K$ and $R \simeq 20\,000~\rm{R_{\odot}}$ (see their table 2).

The work of Shara et al. (2010a) extends the previous nova models of Yaron et al. (2005).
The data of six of their models is given in their tables 2--4.
In the new extreme nova models of Shara et al. (2010a) the mass of the white dwarf is $0.4$~--~$0.65~\rm{M_{\odot}}$
and the accretion rate is as low as $10^{-12}$~--~$10^{-10}~\rm{M_{\odot} \yr^{-1}}$.
The ejected mass in the novae ranges between $5.3 \times 10^{-4}$~--~$ 2.2 \times 10^{-3}~\rm{M_{\odot}}$
at velocities of $150$~--~$480 \km \s^{-1}$.

Shara et al. (2010a) claim that the eruption of M85~OT2006 was a nova,
and suggest that it can be explained in the frame of the new extreme nova models.
Their conclusion is mainly based on the following.
(1) Their new results showing that novae can reach comparable luminosities of ILOTs such as
M85~OT2006 and M31~RV (few $\times 10^7~\rm{L_{\odot}}$).
(2) The models produce red eruptions, as the spectra of ILOTs.

We hereby show that the mass ejected in the eruption of M85~OT2006 is much larger than the nova models can produce.
Let us consider the M85~OT2006 transient according to data given by Rau et al. (2007), discussed above.
The column density required above the photosphere is given by
\begin{equation}
\Delta N \simeq \rho \Delta r = \frac{2}{3} \frac{1}{\kappa} =0.67 \g \cm^{-2} \left(\frac{\kappa}{1 \cm^2 \g^{-1}}\right)^{-1}
\label{eq:rho_deltar}
\end{equation}
where $\rho$ is the average density above the photosphere, $\Delta r$ is the thickness of the shell above the photosphere,
and $\kappa$ is its average opacity.

We parameterize the thickness of the shell with $\Delta r = \beta R$ with $\beta \sim 0.1$.
The final mass we obtain for the shell above the photosphere changes by a factor $<2$ for $0.01 \leqslant \beta \leqslant 0.1$.
The reason is because the value of $\beta$ determines the density that weakly influences the opacity.
We use opacities from Ferguson et al. (2005) (slightly extrapolated), using
compositions from Asplund et al. (2004) with hydrogen abundance $\rm{X}=0.7$ and metallicity $\rm{Z}=0.1$;
other composition from Lodders (2003) gives very close opacity values.
The total mass above the photosphere is
\begin{equation}
\begin{split}
M_{\rm{ph}} &= 4 \pi R^2 \rho \Delta r \\
&\simeq 2 \times 10^{-3} \left(\frac{R}{10^4~\rm{R_{\odot}}}\right)^2 \left(\frac{\kappa}{1 \cm^2 \g^{-1}}\right)^{-1}~\rm{M_{\odot}}.
\end{split}
\label{eq:M_ph}
\end{equation}

Using the temperature at peak luminosity we find that the opacity at the peak luminosity is $\kappa \simeq 1.3 \times 10^{-3}$.
Substituting the opacity and the radius at peak luminosity in equation (\ref{eq:M_ph})
we find that the mass above the photosphere at the time of peak luminosity is $M_{\rm{ph}} \sim 0.2~\rm{M_{\odot}}$.
Our results are summarized in Table \ref{tab:M_ph}.
The real amount of mass is even larger, as some mass is well above the photosphere and we expect a large amount of mass to be
below the photosphere as well.
The largest value of ejected mass in the nova models of Shara et al. (2010a) is $ 2.2 \times 10^{-3}~\rm{M_{\odot}}$ or
two orders of magnitude below the expected ejected mass we calculate.
We therefore conclude that M85~OT2006 is not a nova.
\begin{table*}
\caption{\footnotesize{
Estimate of the mass above the photosphere in the eruption of M85~OT2006.
Data for effective temperature and photospheric radius are taken from Rau et al. (2007).
Opacities are from Ferguson et al. (2005) (slightly extrapolated), using
the composition from Asplund et al. (2004) with Hydrogen abundance $\rm{X=0.7}$ and metallicity $\rm{Z}=0.1$.
The estimated mass above the photosphere is calculated from equation (\ref{eq:M_ph}).
Results are given for two values of the relative thickness of the photosphere $\beta = \Delta r / R$.
}}
\label{tab:M_ph} \medskip
\begin{center}
\footnotesize{
\begin{tabular}{l l l l l l l}
\hline
\multicolumn{1}{c}{Time} & \multicolumn{1}{c}{$T_{\rm{eff}}~[\K]$}   & \multicolumn{1}{c}{$R[~\rm{R_{\odot}}]$}     & \multicolumn{2}{c}{$\kappa [\cm^2 \g^{-1}]$}   & \multicolumn{2}{c}{$M_{\rm{ph}}~[\rm{M_{\odot}}]$}\\
\multicolumn{1}{c}{}     & \multicolumn{1}{c}{}     & \multicolumn{1}{c}{} & \multicolumn{1}{c}{$\beta=0.1$}      & \multicolumn{1}{c}{$\beta=1$} & \multicolumn{1}{c}{$\beta=0.1$}      & \multicolumn{1}{c}{$\beta=1$}\\
\hline
Peak                & $4600$    & $3600$    & $1.3 \times 10^{-3}$ &$1.0 \times 10^{-3}$ & 0.20                 & $0.26$               \\
Late                & $950$     & $20\,000$ & $4.45$               & $2.35$              & $1.8 \times 10^{-3}$ & $3.5 \times 10^{-3}$ \\
\hline
\end{tabular}
}
\end{center}
\end{table*}
\begin{figure}[t]
\resizebox{0.5\textwidth}{!}{\includegraphics{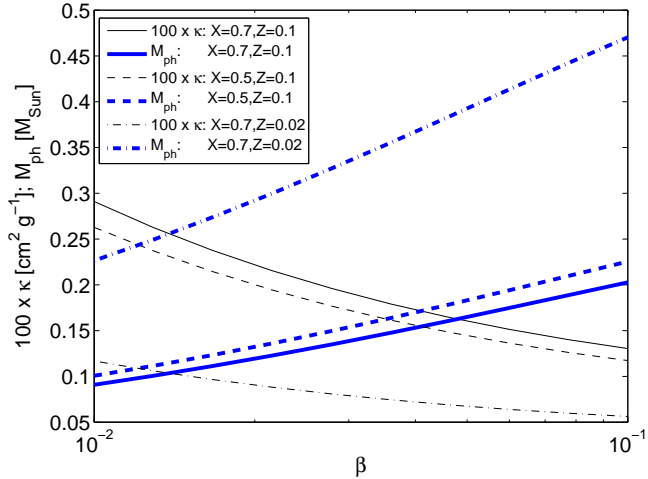}}
\caption{\footnotesize{The photospheric mass and its opacity expected from M85~OT2006 at (or right after) peak luminosity,
when effective temperature is $T_{\rm{eff}} \simeq 4600\K$ and the stellar radius is $R \simeq 3600~\rm{R_{\odot}}$,
plotted as a function of the relative thickness of the photosphere $\beta = \Delta r / R$.
Different cases are for different values of Hydrogen abundance X and metallicity Z (see legend).
}}
\label{fig:Mph_kappa}
\end{figure}

Previously in KFS10, we used the assumption of Ofek et al. (2008) that the total ejected mass of M85~OT2006 is $0.1~\rm{M_{\odot}}$,
and consequently obtained a total energy of $\sim 1.4 \times 10^{47} \erg$.
Our new estimate of ejected mass is much higher and we update our estimate of the total energy in M85~OT2006.
Our new estimate, taking ejected mass velocity of $\sim 870 \km \s^{-1}$ (Rau et al. 2007)
is $1.6 \times 10^{48}$~--~$4.6 \times 10^{48}\erg$, corresponding to $0.2~\rm{M_{\odot}}$ and $0.6~\rm{M_{\odot}}$, respectively.
These values are summarized in Table \ref{tab:data} and the updated location of M85~OT2006 is shown in the ETD (Figure \ref{fig:totEvst}).

We calculate the total (radiated and kinetic) energy of the new nova models of Shara el al. (2010a)
and plot them in Figure \ref{fig:totEvst}.
We include all models from Yaron et al. (2005) (red crosses) including those we omitted in KFS10.
The data of Shara et al. (2010a) is missing the parameter $t_{\rm{ml}}$, the duration of the mass-loss phase,
which was given in the previous list of Yaron et al. (2005).
The six extreme nova models of Shara et al. (2010a) are therefore plotted twice in Figure \ref{fig:totEvst}.
First, taking $t_{3,\rm{bol}}$ instead of $t_{\rm{ml}}$ in the calculation for the total energy (red open diamonds).
Second, following rough average values from Yaron et al. (2005), that show the relation $t_{\rm{ml}}\simeq 0.1 t_{3,\rm{bol}}$
(red filled diamonds).
We find the new extreme nova models to be located in the lower, less populated part of the OTS.
This implies that novae are very unlikely to account for the ILOTs.

\subsection{M31~RV: Seeing an Accretion Disk Through the Ejecta}
\label{subsec:Known_Transients_M31}

The outburst of M31~RV occurred in 1988 and had a peak luminosity of $\sim 8 \times 10^5~\rm{L_{\odot}}$,
declining on a timescale of $\sim 70 \days$ (Rich et al. 1989; Mould et al. 1990).
The progenitor is located in the central bulge of M31 (Rich et al. 1989), where the stellar population is old.
The ejecta was estimated to have a mass of $0.001$~--~$0.1~\rm{M_{\odot}}$ and an expansion velocity of
$100$~--~$500 \km \s^{-1}$ (Mould et al. 1990).
About $54$ days after the peak the spectra of the ILOT was compatible with a blackbody temperature of $\sim 2000 \K$,
and $79$ days after the peak it dropped to $\sim 1050 \K$ (Mould et al. 1990).
Radiation from a $\sim 1000 \K$ dust shell with a radius of $\sim 8000~\rm{R_{\odot}}$ was observed $\sim 70 \days$ after the eruption,
inferring that the average expansion velocity over this period was $\sim 920 \km \s^{-1}$.

Based on some similarities with V838~Mon, Soker \& Tylenda (2003; Tylenda \& Soker 2006)
suggested that M31~RV eruption was caused by mergeburst.
According to the theme of the present paper, a merging processes can be replaced with a rapid
mass transfer episode in a binary system.
In a recent paper, Shara et al. (2010b) analyzed HST archival data, and found that in 1994
the star's luminosity was $\sim 10^3~\rm{L_{\odot}}$ and its temperature was $> 40\,000 \K$.
Shara et al. (2010b) also presented new HST observations of the surroundings of M31~RV from 2008,
and claimed to detect an energy source with an effective temperature of $\sim 8000 \K$ at the location of the progenitor.
According to Shara et al. (2010b), there is currently no nova model that can produce the observed peak luminosity of M31~RV.
However, they suggested that with a usage of the correct opacity (that presently their model lacks)
low mass white dwarfs accreting at low rates can in principle
produce low temperature and very luminous novae, as required in the case of M31~RV.

Motivated by the new HST observations of Shara et al. (2010b) we consider transients with low ejected mass.
Located among old stars, the mass of M31~RV is expected to be low.
Possibly, the ILOT is a merger event of two low mass stars or a mass transfer episode to a low mass star.
In such cases the mass transferred to the accreting star is low, and the ejected mass is even lower.
We suggest that the ejected mass was small, and quickly became optically thin.
Part of the ejected material might have not reached the escape velocity, and fell back toward the star.

In addition, in our model the inflated envelope has also contained very little mass ($< 0.01~\rm{M_{\odot}}$).
This is different than the situation in V838~Mon, where the inflated envelope
was more massive, and is long-lived (Tylenda, Kamiski \& Schmidt 2009).
In the case of V838~Mon $\sim 0.1$~--~$0.3~\rm{M_{\odot}}$ has been accreted on a radiative
envelope of a B-star (Tylenda, Soker \& Szczerba 2005; Tylenda \& Soker 2006).
On the other hand, in the case of M31~RV, the stellar envelope is convective,
and as we show below that only $\sim 0.05~\rm{M_{\odot}}$ has been accreted onto it.
The convective envelope has quickly adjusted itself to the accreted mass.
A large envelope with a very low mass cannot support itself
(as is known for AGB stars for example; Soker 1992), and therefore the inflated envelope
in the case of M31~RV has collapsed onto the low mass star.

The gas in the inflated envelope that later collapsed, and the gas that did not reach the
escape velocity at larger distances, fell back toward the star.
In case of a surviving companion, the companion could have continued to
transfer mass to the mass accreting star.
In any case, instead of a long-lived inflated envelope as in the case of V838~Mon,
an accretion disk was formed after about a year.

We can propose some typical numbers, but with very large uncertainties.
The fall back mass was $M_{\rm{fb}} \sim 10^{-3}~\rm{M_{\odot}}$.
To account for a luminosity of $\sim 1000~\rm{L_{\odot}}$ after several years (Shara et al. 2010b) an
accretion rate of $\sim 10^{-5}~\rm{M_{\odot}} \yr^{-1}$ is required.
The disk can live for $\sim 100 $ years.

Because of the high specific angular momentum in the binary system the fall back
gas formed an accretion disk around the accreting star.
After a relatively short time ($\sim 1 \yr$ in the case of M31~RV)
one would observe the central star and its accretion disk rather than a huge inflated envelope.
The accretion disk and its boundary layer become the dominant illuminating source.

We start by following the calculation of Shore \& King (1986), who studied the physics
of accretion disk boundary layers.
For a star with mass $M_a$ and radius $R_a$, accreting at a rate of $\dot{M_a}$,
the luminosity of the boundary layer would be half of the gravitational luminosity
\begin{equation}
\begin{split}
L_{\rm{BL}} &= \frac{1}{2} \frac{GM_a\dot{M_a}}{R_a} \\
&\simeq 150 \left(\frac{M_a}{1~\rm{M_{\odot}}}\right)
\left(\frac{\dot{M_a}}{10^{-5}~\rm{M_{\odot}}~\yr^{-1}}\right)
\left(\frac{R_a}{1~\rm{R_{\odot}}}\right)^{-1}
~\rm{L_{\odot}}. 
\end{split}
\label{eq:L_BL}
\end{equation}
Having an optically thick boundary layer with a thickness $\delta_{\rm{BL}}$,
it will heat to a temperature of
\begin{equation}
\begin{split}
T_{\rm{BL}} &= \left(\frac{L_{\rm{BL}}}{4 \pi \sigma R_a \delta_{\rm{BL}}}\right)^{\frac{1}{4}} \\
&\simeq 3.3 \times 10^4 \left(\frac{M_a}{1~\rm{M_{\odot}}}\right)^{\frac{1}{4}}
\left(\frac{\dot{M_a}}{10^{-5}~\rm{M_{\odot}}~\yr^{-1}}\right)^{\frac{1}{4}}
\left(\frac{R_a}{1~\rm{R_{\odot}}}\right)^{-\frac{1}{2}} \\
&\left(\frac{\delta_{\rm{BL}}}{10^{10}\cm}\right)^{-\frac{1}{4}}
\K, 
\end{split}
\label{eq:T_BL}
\end{equation}
where $\sigma$ is the Stefan-Boltzmann constant.

The temperature of the boundary layer for the parameters used in equations (\ref{eq:L_BL}) and (\ref{eq:T_BL})
is compatible with the hot temperatures of M31~RV as seen in the observations of Shara et al. (2010b) at late times.
The accretion luminosity is evenly divided between the boundary layer and the accretion disk itself.
The accretion disk temperature is much lower than the temperature of the boundary layer.
Hence, it is expected that the averaged observed temperature will be lower than the temperature of the boundary layer.

Let us now check what would be the optical depth of the transient a few years after its eruption.
We will assume spherical symmetry.
The thickness of the shell, $\Delta r_{\rm{er}}$, ejected during the eruption is determined from the timescale of the eruption $\Delta t_{\rm{er}}$
in which the material is ejected, and the velocity $v_{\rm{er}}$ it had during the eruption
\begin{equation}
\Delta r_{\rm{er}} \simeq 20 \left(\frac{v_{\rm{er}}}{500 \km \s^{-1}}\right)
\left(\frac{\Delta t_{\rm{er}}}{70 \days}\right) \AU. 
\label{eq:delta_r_er}
\end{equation}
We take the mass of the spherically expanding shell $M_{\rm{ej}}$, and its average expansion velocity $v_{\rm{ej}}$.
After a time period $\Delta t$ it would reach a distance of
\begin{equation}
r_{\rm{ej}} \simeq 1200 \left(\frac{v_{\rm{ej}}}{920 \km \s^{-1}}\right)
\left(\frac{\Delta t}{6 \yr}\right) \AU, 
\label{eq:r_ej}
\end{equation}
where the calibration is from M31~RV and the latest HST observations of Shara et al. (2010b).
The average density in the shell is therefore
\begin{equation}
\begin{split}
\rho_{\rm{ej}} &= \frac{M_{\rm{ej}}}{4 \pi r_{\rm{ej}}^2 \Delta r_{\rm{er}}} \\
&\simeq 1.7 \times 10^{-17} \left(\frac{M_{\rm{ej}}}{0.01~\rm{M_{\odot}} }\right)
\left(\frac{r_{\rm{ej}}}{1200 \AU}\right)^{-2} \\
&\left(\frac{\Delta r_{\rm{er}}}{20 \AU}\right)^{-1} \g \cm^{-3}, 
\end{split}
\label{eq:rho_ej}
\end{equation}
and the column density of the shell is
\begin{equation}
\begin{split}
\Delta N_{\rm{ej}} &\simeq \rho_{\rm{ej}} \Delta r_{\rm{er}} \\
&\simeq 5.2 \times 10^{-3} \left(\frac{M_{\rm{ej}}}{0.01~\rm{M_{\odot}} }\right)
\left(\frac{r_{\rm{ej}}}{1200 \AU}\right)^2 \g \cm^{-2}. 
\end{split}
\label{eq:delta_N_ej}
\end{equation}

For a shell temperature of $1000 \K$ with Hydrogen abundance $\rm{X}=0.7$ and metallicity $\rm{Z}=0.02$ the opacity
(from the same data by Ferguson described above) is $\kappa_{\rm{ej}}= 0.03 \cm^2 \g^{-1}$, 
and the optical depth is
\begin{equation}
\begin{split}
\tau_{\rm{ej}} &\simeq \kappa_{\rm{ej}} \Delta N_{\rm{ej}} \\
&\simeq 1.6 \times 10^{-4} \left(\frac{\Delta N_{\rm{ej}}}{5.2 \times 10^{-3} \g \cm^{-2}}\right)
\left(\frac{\kappa_{\rm{ej}}}{0.03 \cm^2 \g^{-1}}\right). 
\end{split}
\label{eq:tau_ej}
\end{equation}
For temperatures in the range $\sim 600$~--~$\sim 3000 \K$ the opacity is $\kappa_{\rm{ej}}<1$ 
and therefore the result that the shell being optically thin is not sensitive to temperature.
If dust is formed the opacity can reach much higher values.
Several researchers have suggested that dust opacities can reach values of $\sim 10 \cm^2 \g^{-1}$ and even higher
(e.g. Helling et al. 2000; Henning \& Stognienko 1996; Pollack \& Mckay 1985; Pollack et al. 1994; Semenov et al. 2003).
However, these high opacities were not obtained for densities as low as $\sim 10^{-17}$, as in our case.
According to the studies discussed above for such densities, the opacity is $\kappa_{\rm{ej}}<1$ in the temperature range $\sim 100$~--~$\sim 1000 \K$,
which is the relevant range for our study.
{}From equations (\ref{eq:r_ej}), (\ref{eq:rho_ej}), (\ref{eq:delta_N_ej}) and (\ref{eq:tau_ej}),
the time period it would take for the dusty shell to become optically thin is
\begin{equation}
\begin{split}
\Delta t &= \frac{1}{v_{\rm{ej}}} \left(\frac{\kappa_{\rm{ej}} M_{\rm{ej}}}{4 \pi \tau_{\rm{ej}}}\right)^{\frac{1}{2}} \\
&\simeq 0.4 \left(\frac{v_{\rm{ej}}}{920 \km \s^{-1}}\right)^{-1}
\left(\frac{\kappa_{\rm{ej}}}{1 \cm^2 \g^{-1}}\right)^{\frac{1}{2}}\\
&\left(\frac{M_{\rm{ej}}}{0.01~\rm{M_{\odot}} }\right)^{\frac{1}{2}}
\left(\frac{\tau_{\rm{ej}}}{1}\right)^{-\frac{1}{2}} \yr. 
\end{split}
\label{eq:delta_t}
\end{equation}
We see that even if dust grains are formed, a few years after the eruption $\tau_{\rm{ej}} << 1$, and the shell becomes optically thin.
This result holds even if the ejected mass is as high as $M_{\rm{ej,max}}=0.1~\rm{M_{\odot}}$.

We therefore propose the following scenario for the outburst of M31~RV.
This scenario with different parameters is relevant to other ILOTs with low mass progenitors as well.
An interaction between two old stars, a primary with a mass of $\sim 1~\rm{M_{\odot}}$
and a less massive companion, led to the ejection of material that created a shell around the primary.
Alternatively, this interaction might have been a mass loss episode of the companion due to tidal interaction with the primary.
Approximately, a mass of $M_{\rm{ej}} \sim 0.01~\rm{M_{\odot}}$ was ejected and created an expanding shell around the stars.
The energy to eject the shell comes from accretion of material onto the star(s).
The kinetic energy of the ejecta is
\begin{equation}
E_{\rm{er,kin}} \simeq 8.4 \times 10^{46}
\left(\frac{M_{\rm{ej}}}{0.01~\rm{M_{\odot}} }\right)
\left(\frac{v_{\rm{ej}}}{920 \km \s^{-1}}\right)^{2} \erg. 
\label{eq:Ekin}
\end{equation}
The radiated energy $E_{\rm{er,rad}} \simeq 10^{46} \erg$ (Mould et al. 1990).
Therefore the total energy of the eruption is
$E_{\rm{er,tot}} = E_{\rm{er,kin}} + E_{\rm{er,rad}} \simeq 10^{47} \erg$. 
We assume that the companion which donated the accreted and ejected material had a highly eccentric orbit $e \gtrsim 0.85$,
and that the accretion episode occurred close to periastron where the tidal force was maximal.
Therefore, the accreted material fell onto the star at approximately the free-fall velocity.
Under this assumption and according to the virial theorem,
the potential energy was $\langle U \rangle = -2 \langle E_{\rm{er,tot}} \rangle = 1.9 \times 10^{47} \erg$. 
The accreted mass needed to supply this amount of energy is
\begin{equation}
\begin{split}
M_{\rm{acc}} &= \frac{2 E_{\rm{er,tot}} R_a}{GM_a} \\
&= 0.05 \left(\frac{E_{\rm{er,tot}}}{10 \times 10^{47} \erg}\right)
\left(\frac{R_a}{1~\rm{R_{\odot}}}\right)
\left(\frac{M_a}{1~\rm{M_{\odot}}}\right)^{-1}
~\rm{M_{\odot}}.
\end{split}
\label{eq:Macc}
\end{equation}
In other words, most of the gravitational energy of the accreted material went to kinetic energy to eject the shell,
and a small part went to radiated energy.

We conclude that after a time period $\Delta t \sim 1 \yr$ the shell became optically thin and the with lack of obscuration from an inflated envelope
the disk became observable (equation \ref{eq:delta_t}).
Therefore, a few years after the eruption a blue energy source is seen at the location of the progenitor,
as observed by Shara et al. (2010b).

\section{SUMMARY}
\label{sec:summary}

In a previous paper (KFS10) we grouped several Intermediate Luminosity Optical Transients (ILOTs),
together with major Luminous Blue Variable (LBV) eruptions.
These objects occupy the Optical Transients Stripe (OTS), between novae and supernovae in the Energy-Time Diagram
(ETD; Figure \ref{fig:totEvst}),
where the transients' total energy (kinetic and radiated; or total available energy when a model exist) is plotted against their decline timescale.

Recent observations of new objects (section \ref{sec:Recent_Transients})allow a better placement the OTS
on the Energy-Time plane.
The LBV eruptions of NGC~3432~OT2 (Pastorello et al. 2010),
bridge the gap between the upper and lower parts of the OTS (see Figure \ref{fig:totEvst}).
NGC~3432~OT serves as  strong support for grouping LBV major eruptions together with ILOTs.
This allows us to overcome a drawback of the original proposal, hence strengthening the usage of the ETD.
The upper bound of the OTS comes from accretion considerations and has a constant luminosity of $\sim 10^{42} \erg ~\rm{s^{-1}}$,
for a maximal accretion rate of $\sim 10 \%$ of the accretor mass (equation \ref{eq:Lmax}).
The lower part of the OTS is only observationally constraint, and has no theoretical limit.
It is however evident that the luminosity of ILOTs is larger than novae despite a small overlap
(which we explain in section \ref{sec:theEnergyTimeDiagram}).

An important contribution to our physical understanding of ILOTs is the new ILOT
V1309~Sco (Nakano 2008) and its similarity to the eruption of V838~Mon,
as noted by Mason et al. (2010) and later confirmed by Tylenda et al. (2011).
The strong similarity in light curves (Figure \ref{fig:V1309Sco_V838Mon_DIM}; after time rescaling)
hints to a basic common physical mechanism.
We fully accept the mergeburst model of V838~Mon and V1309~Sco, and we are not proposing an alternative.
On the contrary, we extend the mergeburst model by considering a massive and short duration mass transfer episode
that do not destroy the mass losing star.
A merger process is considered to be an extreme case of mass transfer events.

We refute claims by Shara et al. (2010a) that the ILOT M85~OT2006 was a nova in light of recent extreme nova models.
We show that the energy of M85~OT2006 is much too high for any nova (even extreme nova),
and place it in the ILOTs part of the ETD (section \ref{subsec:Known_Transients_M85}).
Another ILOT, M31~RV, was observed in an old stellar population.
This implies that the processes which lead to its eruption involved low mass stars.
In our accretion model, a relatively small amount of mass was ejected from the system,
and a small amount of mass was in the inflated envelope.
The small amount of ejected mass ensured that the shell became transparent $\sim 1 \yr$ after eruption.
In addition, with little mass in the inflated envelope, the envelope collapsed within few years.
The fall-back material in our binary model has high specific angular momentum, such that
an accretion disk is formed around the star.
The accretion disk and its boundary layer dominate the radiation.
The temperature is relatively high, $\sim 5000$~--~$50\,000 \K$.
This explains the blue-ward evolution of M31~RV years after eruptions (section \ref{subsec:Known_Transients_M31}), as observed by Shara et al. (2010b).

More transient objects will very likely be observed in coming years.
The OTS is a strong tool in analyzing these transients, as well as light curve matching as we demonstrate here.
Other research groups have also begun to adopt energy considerations.
For example, the .Ia supernova SN2010X (Kasliwal et al. 2010b) was analyzed in terms of its
total energy, rather than its peak magnitude, as we suggested in KFS10.
We encourage other groups to do the same, and to check whether the transient events they observe are located on the OTS.

We thank Avishay Gal-Yam, Nathan Smith and Romek Tylenda for helpful comments.
We especially thank Stephen Rafter for most helpful comments.
We acknowledge Mansi M. Kasliwal for providing us the R-band observations of PTF10fqs.
This research was supported by the Asher Fund for Space Research at the Technion
and a grant from the Israel Science Foundation.


\begin{references}
\footnotesize{

\reference{} Asplund, M., Grevesse, N., Sauval, A. J., Allende Prieto, C. \& Kiselman, D. 2004 A\&A, 417, 751

\reference{} Bath, G. T. \& Pringle, J. E. 1982, MNRAS, 199, 267

\reference{} Berger, E., et al. 2009, ApJ, 699, 1850

\reference{} Damineli, A. 1996, ApJ, 460L, 49

\reference{} Dubus, G., Hameury, J.-M., \& Lasota, J.-P. 2001, A\&A, 373, 251

\reference{} Frew, D. J. 2004, JAD, 10, 6

\reference{} Ferguson, J. W., Alexander, D. R., Allard, F. et al. 2005, ApJ, 623, 585 

\reference{} Henning Ch. \& Stognienko, R. 1996, A\&A, 311, 291 

\reference{} Helling, Ch., Winters, J. M.,\& Sedlmayr, E. 2000, A\&A, 358, 651

\reference{} Kashi, A. 2010, MNRAS, 405, 1924

\reference{} Kashi, A., Frankowski, A. \& Soker, N., 2010, ApJ, 709L, 11 (KFS10)

\reference{} Kashi, A. \& Soker, N. 2010, ApJ, 723, 602

\reference{} Kasliwal, M. M. et al. 2010a, preperint (arXiv:1005.1455)

\reference{} Kasliwal, M. M. et al. 2010b, ApJ, 723L, 98

\reference{} Kenyon, S. J., Gallagher, J. S., III 1985, ApJ, 290, 542

\reference{} Kenyon, S. J., \& Webbink, R. F. 1984, ApJ, 279, 252

\reference{} Kulkarni, S. R. et al. 2007a, Nature, 447, 458

\reference{} Kulkarni, S. R. et al. 2007b, Nature, 449, 1

\reference{} Lodders, K. 2003, ApJ, 591, 1220

\reference{} Mason, E., Diaz, M., Williams, R. E., Preston, G. \& Bensby, T. 2010, A\&A, 516A, 108

\reference{} Mould et al. 1990, ApJ, 353, 35

\reference{} Nakano, S. 2008, IAUC, 8972

\reference{} Ofek, E. O. et al. 2008, ApJ, 674, 447

\reference{} Puckett, T. \& Gauthier, S. 2002, IAUC, 7863, 1

\reference{} Pastorello et al. 2010, MNRAS, 408, 181

\reference{} Pollack, J. B., Hollenbach, D., Beckwith, S., Simonelli, D. P., Roush, T., \& Fong, W. 1994, ApJ, 421, 615

\reference{} Pollack, J. B. \& Mckay, C. P. 1985, ICar, 64, 471

\reference{} Rau, A., Kulkarni, S.R., Ofek, E.O. \& Yan, L. 2007, ApJ, 659, 1536

\reference{} Rich, R., Mould, J., Picard, A., Froger, J. \& Davies, R. 1989, ApJL, 341, L51

\reference{} Semenov, D., Henning, Th., Helling, Ch., Ilgner, M. \& Sedlmayr, E. 2003, A\&A, 410, 611

\reference{} Shara, M. M., Yaron, O., Prialnik, D., Kovetz, A. \& Zurek, D. 2010a ApJ, 725, 831 

\reference{} Shara, M. M., Zurek, D., Prialnik, D., Yaron, O. \& Kovetz, A. 2010b ApJ, 725, 824 

\reference{} Shore, S. N. \& King, A. R. 1986, A\&A, 154, 263

\reference{} Smith, N., et al. 2009, ApJL, 697L, 49

\reference{} Smith, N. 2011, preperint (arXiv:1010.3770)

\reference{} Smith, N. \& Frew, D. 2011, preperint (arXiv:1010.3719)

\reference{} Smith, N., Li, W., Silverman, J. M., Ganeshalingam, M. \& Filippenko, A. V. 2011, preperint (arXiv:1010.3718)

\reference{} Soker, N. 1992, ApJ, 389, 628

\reference{} Soker, N. \& Tylenda, R. 2003, ApJ, 582, L105

\reference{} Soker, N. \& Tylenda, R. 2006, MNRAS, 373, 733

\reference{} Soker, N. \& Tylenda, R. 2007, ASPC, 363, 280

\reference{} Sparks, W. B. et al. 2008, AJ, 135, 605

\reference{} Thompson, T. A., Prieto, J. L., Stanek, K. Z., Kistler, M. D., Beacom, J. F., Kochanek, C. S. 2009, ApJ, 705, 1364

\reference{} Tylenda, R. 2005, A\&A, 436, 1009

\reference{} Tylenda, R., Kamiski, T. \& Schmidt, M. 2009, A\&A, 503, 899

\reference{} Tylenda, R. \& Soker, N. 2006, A\&A, 451, 223

\reference{} Tylenda, R., Soker, N. \& Szczerba, R. 2005, A\&A, 441, 1099

\reference{} Tylenda, R. et al. 2011, preperint (arXiv:1012.0163)

\reference{} Wagner, R. M. et al. 4004, PASP, 116, 326

\reference{} Yaron, O., Prialnik, D., Shara, M. M., Kovetz, A. 2005, ApJ, 623, 398

}

\end{references}
\end{document}